# The light curve of T Cephei: humps, long-term changes and recent phenomena

Mike Goldsmith[1]

*Independent researcher*

1 June 2026

**Abstract**

The long-period Mira variable T Cephei has been observed for nearly 150 years, during which more than 130 cycles have been recorded. Using visual observations from the AAVSO database, maxima, minima, pulsation periods, amplitudes and hump characteristics were estimated for 136 cycles from the late nineteenth century to 2026.

Most cycles display a temporary slowing of the rise to maximum light ("humps"). These humps show significant cycle-to-cycle persistence in occurrence, duration and brightness, and are more likely to occur during high-amplitude cycles. Hump onset magnitudes exhibit a lower limit near V ~ 9.3, with many humps beginning near V ~ 8. Cycles with larger amplitudes/deeper minima tend to have humps which are dimmer and more gradual.

Comparison with published spectroscopy supports the interpretation that the humps are associated with changes in TiO-dominated molecular opacity within the stellar atmosphere. A qualitative model involving secondary pressure waves traversing TiO-rich layers is proposed to account for the observed hump morphology and its variability.

The most recent cycles of T Cep have shown unprecedented behaviour. Pulsation amplitudes have fallen to historically low values, the maximum-to-maximum period has shortened substantially, and the light curve has become increasingly asymmetric. In Cycle 139, a substantial hump appeared on the descending branch, a phenomenon never before observed in T Cep. Similarities with the Mira variable T UMi raise the possibility that these recent changes may be due to evolutionary developments.

## Introduction

T Cephei was discovered to be a variable in 1879 by Vitold Ceraski (Ceraski, 1879). AAVSO-collated photometric observations began in 1881 and became plentiful from 1906.

[1] contact: mikegoldsmith639@gmail.com

Like all Mira variables, T Cep's maxima and minima vary somewhat in both magnitude and period. In addition, the majority of cycles exhibit a slowing-down of the increase of light to maximum. This feature is usually called a hump and ranges considerably in duration and morphology. It has been present since the earliest photometric measurements (Figure 1).

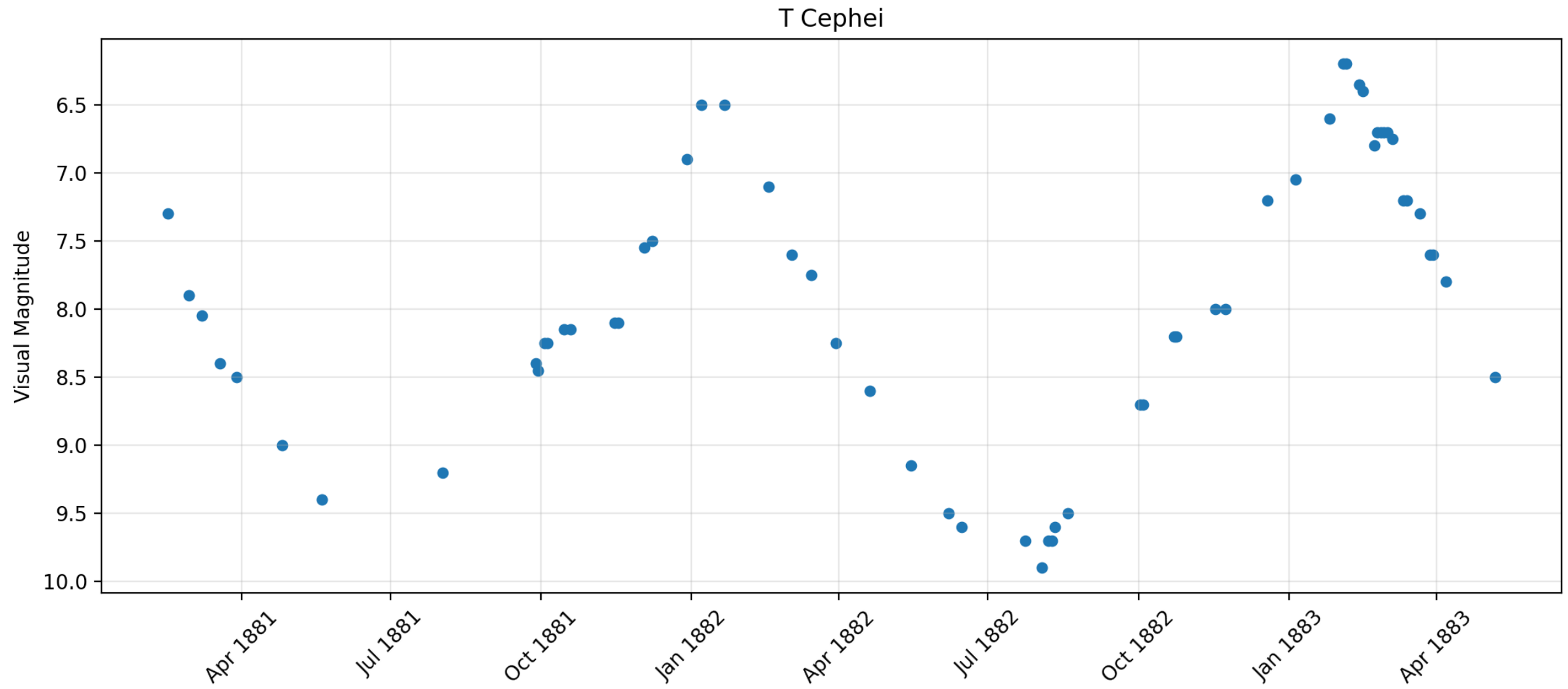

**Figure 1. Earliest AAVSO observations of T Cep, showing hump in the winter of 1881/1882**

Maurice Gavin's spectra of T Cep, two of which were taken close to the start and near the end the 1999 hump, show that TiO-dominated spectral regions remained comparatively unchanged while other features brightened (Gavin 1999; see Figure 2). This is consistent with the accepted model for Mira variability, in which the formation and dissociation of TiO molecules is the major contributor to the change of visible light levels.

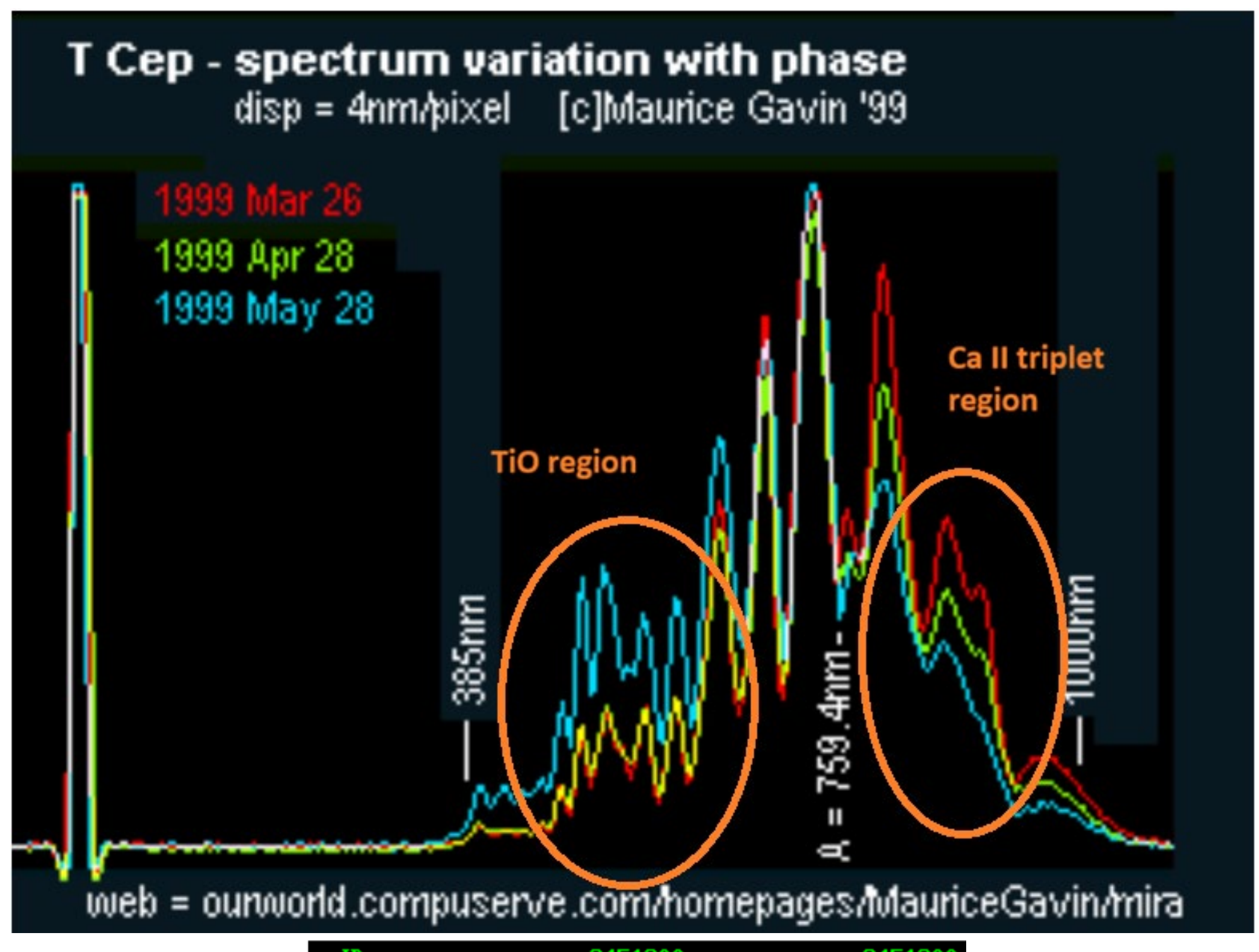

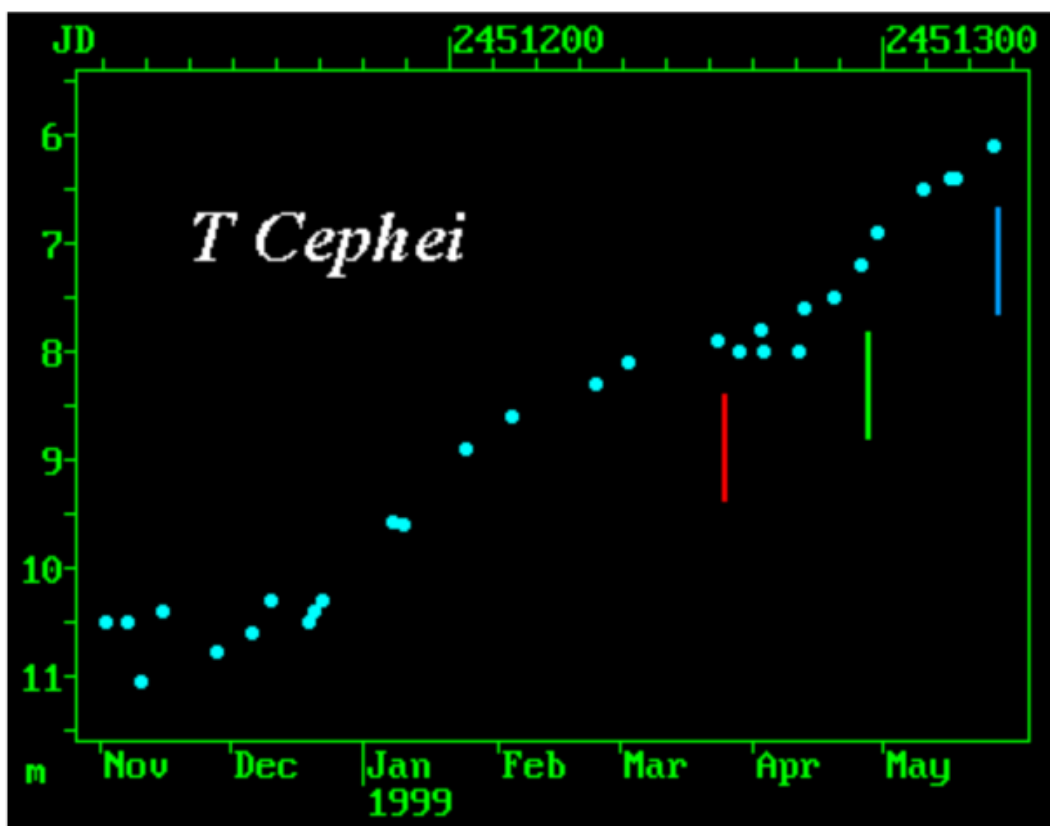

**Figure 2. Figures reproduced from Gavin (1999), courtesy of the British Astronomical Association Variable Star Section. The traces which appear yellow on the upper plot show where the spectra taken near the beginning and end of the hump coincide. The dates when the spectra were taken are indicated by the colour-coded lines in the lower plot. Orange annotations are additions by the present author and do not appear in the original figure.**

Conversely Boyd (2023), which includes plots of spectra taken between December 2021 and December 2022 (JD 2459935 to JD 2460306) over the whole of Cycle 137, does not appear to show a comparable interval of unchanging TiO-band intensity, and this corresponds to the lack of a hump in that cycle (Figure 3).

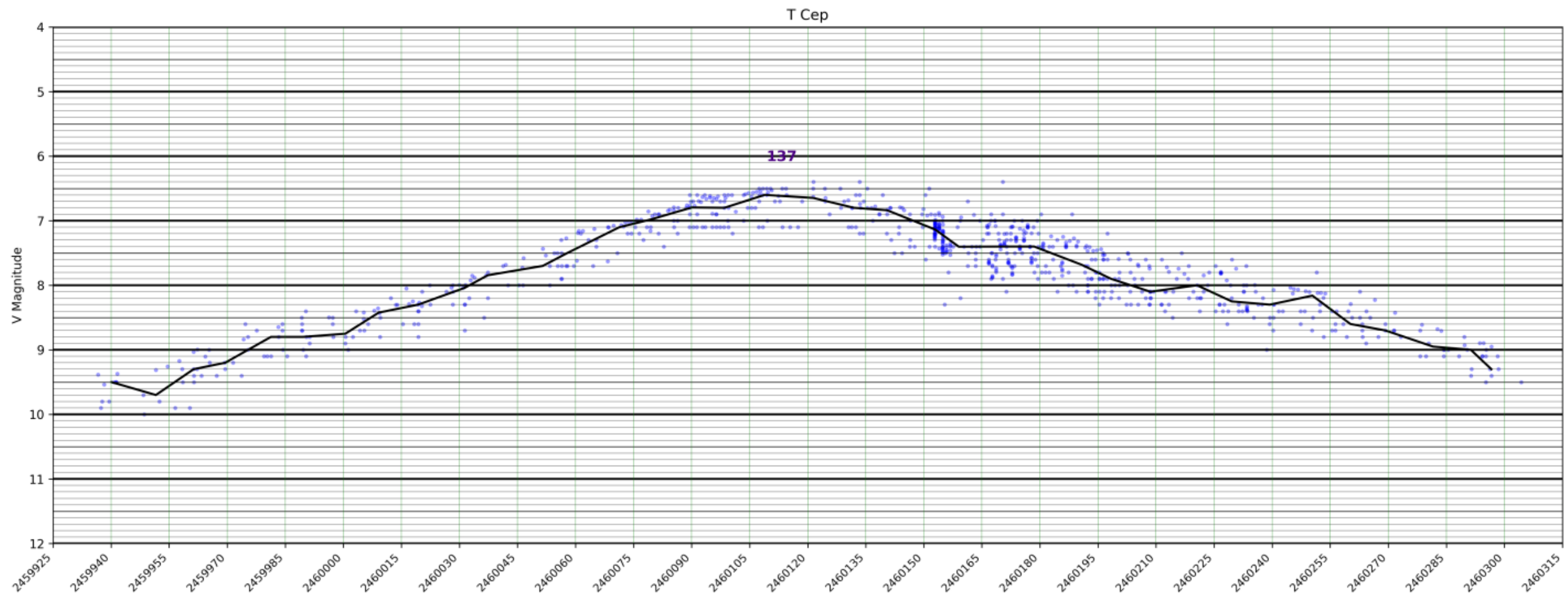

**Figure 3. December 2021 to December 2022 (AAVSO data)**

Such humps are present in several other Mira stars, almost always on the ascending branch (see, e.g. Templeton & Willson, 2004).

The statistical properties of Mira light curves have been the subject of investigation for many decades. Studies have examined long-term variations in period and amplitude (Percy & Colivas 1999; Templeton & Willson 2004), the morphology and asymmetry of Mira light curves (Vardya 1988; Lebzelter 2011), and the occurrence of humps and other secondary features which may be associated with atmospheric shock phenomena (Kudashkina & Rudnitskij 1994). More recently, Merchan-Benitez et al. (2023) analysed asymmetries and long-term changes in Mira light curves using modern datasets.

The present study analyses the statistical properties of the visual light curve of T Cep over approximately 145 years of observations. Particular attention is given to the occurrence and characteristics of humps, including their brightnesses, durations, slopes, cycle-to-cycle persistence, and relationship to pulsation amplitude and period. The recent emergence of unusual light-curve behaviour, including hump formation on the descending branch, is also investigated.

## Method

AAVSO data was analysed to establish the magnitudes and dates of maxima and minima and to estimate the parameters of the humps.

All visual observations were first grouped into 10-day bins. Within each bin, the median Julian Date (JD) and median visual magnitude were calculated. Prior to binning, outliers were rejected using a median/Median Absolute Deviation (MAD) criterion: for bins containing at least five observations, any points differing from the median by more than 3×MAD were excluded.

For each cycle[2], approximate maxima and minima were first identified from the binned light curve. Then a quadratic curve was fitted to several points surrounding the approximate extremum, and the vertex adopted as the cycle maximum or minimum when it lay within the fitted interval; otherwise the brightest or faintest observed point was used directly.

To identify the onset of the hump, sliding slopes were calculated along the light curve. Each slope was obtained from a least-squares linear fit to five consecutive binned points. The steepest slope was selected, and candidate hump starts were then defined as locations where the slope reduced after this point. From these candidates, the location showing the largest reduction in slope was selected as the hump start.

The hump end was determined by first calculating a "post-hump slope" (PPS) from a linear fit to several points immediately preceding the maximum. The ascending branch was then scanned backwards point-by-point. The first slope exceeding 0.75×PPS was identified, and the following point was adopted as the hump end. (The 0.75 figure was obtained by trial and error).

Because of the irregularity of the light curve and the wide variation in hump morphology it is not always clear if a hump is present. Also, the onsets and offsets of some humps are sometimes very gradual, so the assignment of hump parameters is to some extent arbitrary. Hence, some of the automatically-located humps were manually rejected or their parameters adjusted slightly.

**Results**

Table 1 shows the estimated positions of key light curve features per cycle.

**Table 1**

| Cycle | Maximum date | Maximum magnitude | Minimum date | Minimum magnitude | Hump start date | Hump start magnitude | Hump end date | Hump end magnitude |
|---|---|---|---|---|---|---|---|---|
| 3 | | | 2408255 | 9.4 | 2408360 | 8.2 | 2408400 | 8.1 |
| 4 | 2408461 | 6.6 | 2408643 | 9.8 | 2408740 | 8.2 | 2408775 | 8.0 |
| 5 | 2408846 | 6.4 | 2409030 | 9.9 | 2409136 | 8.3 | 2409166 | 8.0 |
| 6 | 2409240 | 6.7 | 2409400 | 9.7 | | | | |
| 7 | 2409627 | 6.5 | 2409794 | 9.5 | 2409880 | 8.2 | 2409890 | 8.3 |
| 8 | 2409978 | 6.6 | 2410229 | 9.4 | 2410280 | 7.8 | 2410320 | 7.6 |

[2] Cycles are numbered relative to the first observation of T Cep made by Ceraski on 31 July 1877. The maximum immediately succeeding this observation (albeit unobserved) is designated the beginning of Cycle 0. Adopting the literature-based mean pulsation period of approximately 388.5 days implies that the first clear AAVSO minimum, at JD 2408255, corresponds to the minimum of Cycle 3.

| Cycle | Maximum date | Maximum magnitude | Minimum date | Minimum magnitude | Hump start date | Hump start magnitude | Hump end date | Hump end magnitude |
|---|---|---|---|---|---|---|---|---|
| 9 | 2410363 | 6.4 | 2410593 | 9.2 | 2410675 | 7.1 | 2410754 | 7.1 |
| 10 | | | 2410948 | 9.7 | 2411055 | 7.9 | 2411100 | 7.8 |
| 11 | 2411155 | 6.2 | 2411330 | 10.1 | | | | |
| 12 | 2411531 | 6.5 | 2411698 | 9.2 | | | | |
| 13 | 2411928 | 6.1 | 2412093 | 10.0 | 2412198 | 7.7 | 2412250 | 7.7 |
| 14 | 2412311 | 6.1 | 2412481 | 9.9 | 2412590 | 8.0 | 2412645 | 7.7 |
| 15 | 2412700 | 5.7 | 2412868 | 10.0 | 2412950 | 8.1 | 2413015 | 7.4 |
| 16 | 2413079 | 5.5 | 2413268 | 10.4 | 2413350 | 8.2 | 2413410 | 8.1 |
| 17 | 2413477 | 6.2 | 2413657 | 10.2 | 2413752 | 8.4 | 2413800 | 8.1 |
| 18 | 2413865 | 6.3 | 2414056 | 10.2 | | | | |
| 19 | 2414229 | 6.1 | 2414423 | 9.3 | | | | |
| 20 | 2414632 | 5.9 | 2414805 | 9.6 | | | | |
| 21 | | | | | | | | |
| 22 | 2415420 | 6.3 | 2415594 | 10.2 | 2415710 | 7.4 | 2415751 | 7.1 |
| 23 | 2415799 | 6.1 | 2416017 | 10.4 | 2416111 | 7.8 | 2416138 | 7.8 |
| 24 | 2416192 | 6.3 | 2416374 | 9.6 | | | | |
| 25 | 2416554 | 5.9 | 2416759 | 9.8 | | | | |
| 26 | 2416957 | 6.1 | 2417178 | 9.4 | | | | |
| 27 | 2417349 | 6.2 | 2417532 | 9.7 | | | | |
| 28 | 2417726 | 6.0 | 2417924 | 9.8 | 2418041 | 6.7 | 2418090 | 6.4 |
| 29 | 2418116 | 6.2 | 2418304 | 9.6 | 2418420 | 6.7 | 2418485 | 6.6 |
| 30 | 2418508 | 6.5 | 2418692 | 9.9 | 2418795 | 6.9 | 2418870 | 6.8 |
| 31 | 2418912 | 6.4 | 2419068 | 10.2 | | | | |
| 32 | 2419287 | 5.9 | 2419458 | 10.6 | 2419585 | 7.9 | 2419620 | 8.0 |
| 33 | 2419677 | 6.0 | 2419842 | 10.0 | 2419973 | 7.0 | 2420013 | 6.9 |
| 34 | 2420064 | 5.9 | 2420235 | 10.1 | 2420350 | 7.0 | 2420390 | 7.0 |
| 35 | 2420454 | 6.1 | 2420626 | 10.3 | 2420709 | 8.2 | 2420780 | 7.6 |
| 36 | 2420835 | 6.2 | 2421017 | 10.4 | | | | |
| 37 | 2421229 | 6.1 | 2421413 | 10.3 | 2421490 | 8.2 | 2421540 | 7.6 |
| 38 | 2421609 | 5.8 | 2421804 | 9.9 | 2421910 | 6.5 | 2421990 | 6.4 |
| 39 | 2422012 | 6.2 | 2422200 | 10.3 | 2422282 | 7.3 | 2422362 | 7.0 |
| 40 | 2422408 | 6.4 | 2422582 | 10.2 | 2422650 | 7.9 | 2422760 | 6.9 |
| 41 | 2422798 | 6.1 | 2422965 | 10.0 | 2423060 | 7.7 | 2423130 | 7.1 |
| 42 | 2423188 | 6.0 | 2423362 | 10.3 | 2423472 | 7.3 | 2423523 | 6.8 |
| 43 | 2423578 | 5.9 | 2423756 | 10.3 | 2423881 | 7.7 | 2423902 | 7.5 |
| 44 | 2423962 | 6.3 | 2424157 | 10.0 | 2424261 | 7.0 | 2424313 | 6.5 |
| 45 | 2424349 | 6.0 | 2424532 | 10.0 | 2424639 | 7.4 | 2424702 | 6.8 |
| 46 | 2424744 | 6.2 | 2424933 | 10.2 | | | | |

| Cycle | Maximum date | Maximum magnitude | Minimum date | Minimum magnitude | Hump start date | Hump start magnitude | Hump end date | Hump end magnitude |
|---|---|---|---|---|---|---|---|---|
| 47 | 2425131 | 6.1 | 2425318 | 10.2 | | | | |
| 48 | 2425541 | 6.1 | 2425730 | 10.2 | | | | |
| 49 | 2425934 | 6.3 | 2426130 | 10.3 | | | | |
| 50 | 2426330 | 6.3 | 2426521 | 10.4 | 2426615 | 8.0 | 2426650 | 8.0 |
| 51 | 2426724 | 6.3 | 2426915 | 10.1 | 2426990 | 7.9 | 2427050 | 7.0 |
| 52 | 2427125 | 6.0 | 2427315 | 10.5 | 2427381 | 8.5 | 2427459 | 7.6 |
| 53 | 2427513 | 6.1 | 2427706 | 10.1 | 2427800 | 8.0 | 2427870 | 7.8 |
| 54 | 2427924 | 5.7 | 2428105 | 10.0 | 2428200 | 8.2 | 2428250 | 7.6 |
| 55 | 2428317 | 5.8 | 2428501 | 10.6 | 2428600 | 8.5 | 2428661 | 8.1 |
| 56 | 2428710 | 5.8 | 2428894 | 10.9 | 2429000 | 8.6 | 2429031 | 8.6 |
| 57 | 2429123 | 5.8 | 2429301 | 10.3 | | | | |
| 58 | 2429511 | 5.7 | 2429700 | 10.2 | 2429789 | 7.6 | 2429844 | 7.5 |
| 59 | 2429903 | 5.8 | 2430093 | 10.5 | 2430190 | 8.4 | 2430230 | 8.3 |
| 60 | 2430306 | 6.3 | 2430490 | 10.7 | | | | |
| 61 | 2430677 | 5.9 | 2430876 | 9.9 | 2430955 | 7.6 | 2431010 | 7.3 |
| 62 | 2431079 | 5.7 | 2431274 | 10.4 | 2431350 | 7.9 | 2431421 | 7.6 |
| 63 | 2431473 | 6.3 | 2431676 | 10.6 | 2431741 | 8.4 | 2431822 | 8.1 |
| 64 | 2431874 | 6.1 | 2432066 | 10.3 | 2432160 | 7.2 | 2432202 | 7.0 |
| 65 | 2432256 | 5.5 | 2432463 | 10.3 | 2432540 | 8.1 | 2432590 | 7.4 |
| 66 | 2432647 | 6.1 | 2432838 | 10.2 | | | | |
| 67 | 2433045 | 6.0 | 2433224 | 10.4 | 2433325 | 7.9 | 2433372 | 7.9 |
| 68 | 2433438 | 6.2 | 2433605 | 10.2 | | | | |
| 69 | 2433826 | 6.0 | 2433999 | 10.4 | 2434085 | 7.9 | 2434162 | 7.5 |
| 70 | 2434201 | 6.2 | 2434377 | 10.0 | | | | |
| 71 | 2434574 | 5.9 | 2434763 | 9.9 | 2434870 | 7.4 | 2434913 | 7.3 |
| 72 | 2434957 | 6.1 | 2435124 | 10.1 | 2435230 | 6.9 | 2435281 | 7.5 |
| 73 | 2435342 | 6.2 | 2435507 | 10.1 | 2435614 | 7.7 | 2435672 | 7.4 |
| 74 | 2435710 | 6.3 | 2435886 | 10.2 | 2436000 | 7.6 | 2436040 | 7.2 |
| 75 | 2436095 | 6.1 | 2436266 | 10.0 | 2436362 | 8.0 | 2436440 | 6.9 |
| 76 | 2436475 | 6.2 | 2436652 | 10.4 | | | | |
| 77 | 2436841 | 6.3 | 2437025 | 9.9 | 2437140 | 6.6 | 2437210 | 6.4 |
| 78 | 2437222 | 6.2 | 2437393 | 9.7 | | | | |
| 79 | 2437575 | 6.0 | 2437760 | 9.8 | | | | |
| 80 | 2437973 | 6.0 | 2438145 | 9.8 | 2438270 | 7.6 | 2438300 | 7.4 |
| 81 | 2438350 | 6.4 | 2438515 | 9.9 | | | | |
| 82 | 2438724 | 6.1 | 2438915 | 10.1 | 2439025 | 7.7 | 2439060 | 7.5 |
| 83 | 2439117 | 6.5 | 2439302 | 9.8 | 2439414 | 7.0 | 2439463 | 6.7 |
| 84 | 2439501 | 6.2 | 2439685 | 10.1 | 2439811 | 7.0 | 2439832 | 6.9 |

| Cycle | Maximum date | Maximum magnitude | Minimum date | Minimum magnitude | Hump start date | Hump start magnitude | Hump end date | Hump end magnitude |
|---|---|---|---|---|---|---|---|---|
| 85 | 2439889 | 6.4 | 2440078 | 9.5 | 2440183 | 6.5 | 2440221 | 6.3 |
| 86 | 2440259 | 6.0 | 2440460 | 9.7 | 2440570 | 6.5 | 2440614 | 6.4 |
| 87 | 2440638 | 6.3 | 2440816 | 9.2 | 2440952 | 6.7 | 2441000 | 6.4 |
| 88 | 2441029 | 5.9 | 2441210 | 9.7 | 2441341 | 7.1 | 2441382 | 7.0 |
| 89 | 2441426 | 6.3 | 2441607 | 10.0 | 2441704 | 7.6 | 2441772 | 7.4 |
| 90 | 2441823 | 6.2 | 2441995 | 10.3 | 2442080 | 8.2 | 2442160 | 7.4 |
| 91 | 2442205 | 6.2 | 2442389 | 9.9 | 2442465 | 7.8 | 2442513 | 7.7 |
| 92 | 2442591 | 5.8 | 2442797 | 10.5 | 2442881 | 8.4 | 2442934 | 7.7 |
| 93 | 2442990 | 6.5 | 2443190 | 9.7 | | | | |
| 94 | 2443357 | 6.3 | 2443566 | 10.1 | 2443650 | 7.7 | 2443700 | 7.6 |
| 95 | 2443784 | 6.2 | 2443968 | 10.0 | 2444030 | 8.2 | 2444112 | 7.9 |
| 96 | 2444171 | 6.1 | 2444362 | 10.6 | 2444453 | 8.4 | 2444523 | 7.8 |
| 97 | 2444579 | 6.3 | 2444758 | 10.7 | 2444830 | 8.8 | 2444903 | 8.0 |
| 98 | 2444970 | 6.2 | 2445159 | 10.7 | 2445232 | 8.6 | 2445300 | 7.8 |
| 99 | 2445357 | 6.2 | 2445554 | 10.1 | 2445665 | 7.2 | 2445694 | 6.8 |
| 100 | 2445746 | 5.9 | 2445946 | 10.2 | 2446019 | 8.2 | 2446082 | 7.5 |
| 101 | 2446150 | 6.1 | 2446341 | 10.3 | 2446445 | 8.0 | 2446490 | 7.8 |
| 102 | 2446550 | 6.1 | 2446726 | 10.1 | 2446851 | 7.6 | 2446891 | 7.4 |
| 103 | 2446938 | 5.6 | 2447135 | 10.6 | 2447230 | 8.6 | 2447293 | 7.9 |
| 104 | 2447342 | 6.0 | 2447563 | 10.3 | | | | |
| 105 | 2447741 | 6.1 | 2447947 | 10.5 | | | | |
| 106 | 2448144 | 6.1 | 2448345 | 10.7 | 2448453 | 8.2 | 2448502 | 7.6 |
| 107 | 2448552 | 5.8 | 2448739 | 10.4 | 2448841 | 7.8 | 2448860 | 7.8 |
| 108 | 2448945 | 5.8 | 2449152 | 10.7 | 2449260 | 8.7 | 2449305 | 8.1 |
| 109 | 2449347 | 5.9 | 2449549 | 10.8 | 2449640 | 9.3 | 2449690 | 8.3 |
| 110 | 2449755 | 6.5 | 2449952 | 10.5 | 2450040 | 7.6 | 2450094 | 7.3 |
| 111 | 2450149 | 5.6 | 2450344 | 10.6 | 2450410 | 8.9 | 2450482 | 8.0 |
| 112 | 2450549 | 6.3 | 2450753 | 10.9 | 2450830 | 8.5 | 2450910 | 7.3 |
| 113 | 2450951 | 6.4 | 2451146 | 10.6 | 2451230 | 8.1 | 2451270 | 7.9 |
| 114 | 2451345 | 5.8 | 2451539 | 10.5 | 2451641 | 8.1 | 2451680 | 8.0 |
| 115 | 2451743 | 6.2 | 2451916 | 10.6 | 2452000 | 8.1 | 2452073 | 7.2 |
| 116 | 2452129 | 5.9 | 2452299 | 10.4 | 2452415 | 8.1 | 2452455 | 7.9 |
| 117 | 2452511 | 6.2 | 2452693 | 10.4 | 2452783 | 8.0 | 2452853 | 7.2 |
| 118 | 2452900 | 5.9 | 2453086 | 10.2 | 2453170 | 8.0 | 2453232 | 7.5 |
| 119 | 2453288 | 5.7 | 2453479 | 10.0 | | | | |
| 120 | 2453670 | 6.0 | 2453863 | 10.0 | 2453941 | 8.6 | 2453982 | 7.9 |
| 121 | 2454050 | 6.0 | 2454258 | 10.0 | 2454372 | 6.7 | 2454404 | 6.4 |
| 122 | 2454440 | 6.0 | 2454641 | 10.3 | 2454750 | 7.1 | 2454800 | 6.8 |

| Cycle | Maximum date | Maximum magnitude | Minimum date | Minimum magnitude | Hump start date | Hump start magnitude | Hump end date | Hump end magnitude |
|---|---|---|---|---|---|---|---|---|
| 123 | 2454848 | 6.2 | 2455025 | 10.8 | 2455113 | 8.2 | 2455183 | 7.8 |
| 124 | 2455228 | 6.1 | 2455412 | 10.2 | 2455490 | 8.4 | 2455540 | 8.2 |
| 125 | 2455599 | 6.0 | 2455780 | 10.0 | 2455870 | 8.2 | 2455914 | 7.6 |
| 126 | 2455977 | 5.9 | 2456156 | 9.6 | 2456280 | 7.0 | 2456350 | 6.5 |
| 127 | 2456377 | 6.4 | 2456530 | 9.7 | 2456641 | 7.2 | 2456712 | 6.8 |
| 128 | 2456756 | 6.1 | 2456918 | 10.3 | 2457002 | 7.9 | 2457092 | 7.7 |
| 129 | 2457137 | 6.4 | 2457284 | 10.1 | 2457351 | 8.1 | 2457462 | 7.3 |
| 130 | 2457506 | 6.0 | 2457663 | 9.9 | 2457729 | 8.4 | 2457830 | 7.7 |
| 131 | 2457886 | 6.7 | 2458031 | 9.9 | 2458160 | 7.4 | 2458182 | 7.4 |
| 132 | 2458269 | 6.6 | 2458405 | 10.0 | 2458510 | 7.9 | 2458573 | 7.6 |
| 133 | 2458629 | 6.4 | 2458791 | 10.0 | 2458890 | 7.6 | 2458952 | 7.4 |
| 134 | 2459002 | 6.5 | 2459165 | 10.0 | 2459285 | 7.1 | 2459331 | 7.1 |
| 135 | 2459376 | 6.5 | 2459551 | 9.9 | | | | |
| 136 | 2459741 | 6.5 | 2459940 | 9.6 | | | | |
| 137 | 2460113 | 6.6 | 2460330 | 9.6 | | | | |
| 138 | 2460464 | 6.8 | 2460713 | 9.6 | | | | |
| 139 | 2460821 | 6.7 | 2461098 | 9.6 | 2450850 | 7.0 | 2450990 | 7.6 |

**Analyses[3]**

Of the 136 cycles above, humps on the ascending branch are clearly present in 103 cases, and may well exist in several others where observational data are scanty (there are also several cycles with suggestively wide maximal regions which may mark the formation of a "hump" at maximum light. See Figure 4 for an example.) No cycles seem to have more than one ascending-branch hump.

[3] The hump of Cycle 139 is not included in the following statistics, partly because its complex morphology makes its parameters hard to estimate, partly because it may be a different type of feature – see discussion section.

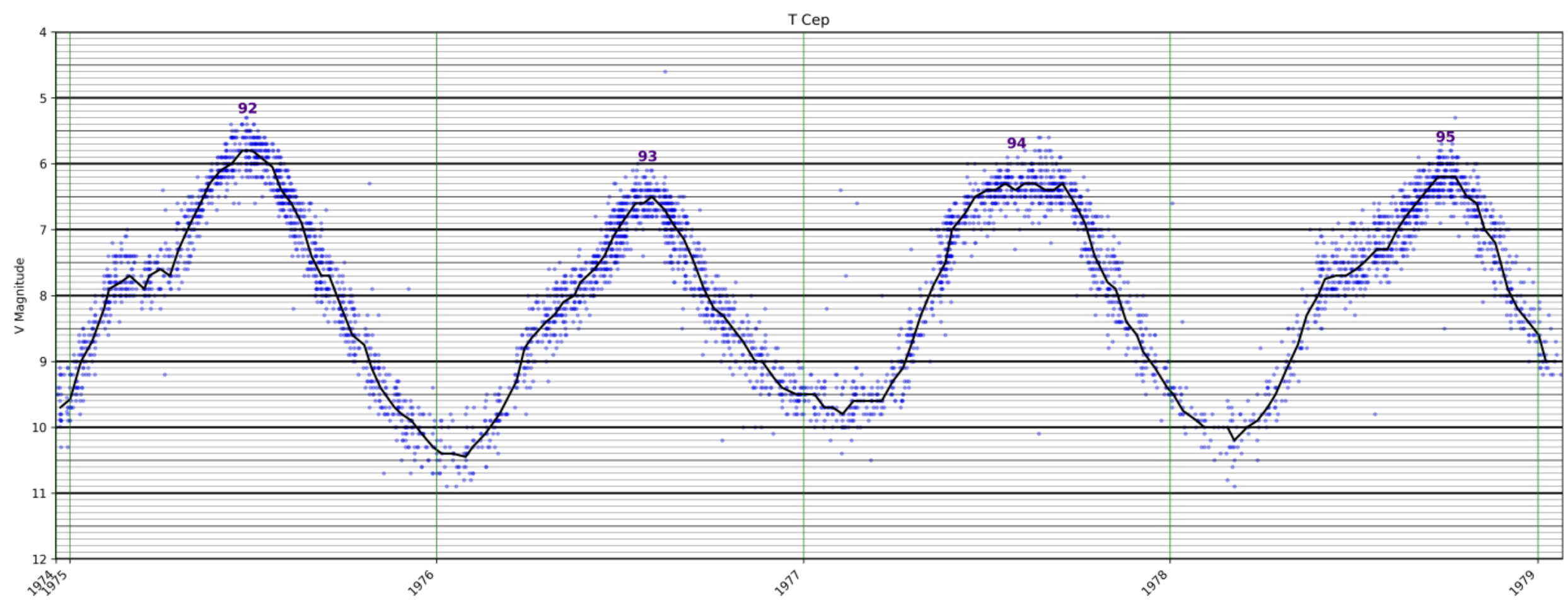

**Figure 4. Possible hump at maximum for Cycle 94**

Consecutive cycles have very similar amplitudes, with an autocorrelation coefficient of 0.75 ($p \approx 8.3\times10^{-25}$).

The mean amplitude of cycles with a hump is 4.03, compared to a mean amplitude of 3.79 when no hump is seen. Applying logistic regression to the data gives $p \approx 0.014$, which means that if amplitude had no effect on hump probability, there would be about a 1.4% chance of obtaining a relationship this strong (or stronger) purely from random sampling variation. So, large-amplitude cycles are significantly more likely to include humps.

Humps also tend to group: their occurrence is not independent from cycle to cycle. A cycle containing a hump is followed by another hump in 81% of cases, compared with only 56% following a hump-free cycle. A 2×2 contingency analysis gives $p \approx 0.004$, indicating statistically significant persistence of hump occurrence from one cycle to the next.

Humps in successive cycles tend to be similar in duration and brightness: the lag-1 autocorrelation coefficients are 0.29 ($p \approx 0.0037$) and 0.38 ($p \approx 1.1\times10^{-4}$) respectively.

The correlation between amplitude and hump duration is negligible, while that between amplitude and hump start brightness is strong: $r \approx -0.44$, $p \approx 4.2\times10^{-6}$.

There is an even stronger correlation between the hump start brightness and the brightness of the minimum $r \approx 0.606$, $p \approx 1.46 \times 10^{-11}$.

Correlation between hump duration and mean hump brightness is negligible.

In other words, while humps in consecutive cycles end to be similar in duration and in brightness, and humps are generally brighter when they follow a bright minimum, it is not the case that bright humps are particularly long-lasting, nor does the duration of the hump seem to relate to the brightness of the preceding minimum.

In 89% of cases, humps end brighter than they begin, and there is a moderate correlation between amplitude and the steepness of the slope of the hump ($r \approx 0.31$, $p \approx 0.0017$), and a much stronger negative correlation between hump start brightness and steepness ($r \approx -0.40$, $p \approx 2.8\times10^{-5}$ (that is, when a hump begins at a relatively early, dim stage of the ascending branch, the slowdown of brightening is relatively slight).

It should be borne in mind, however, that the values for hump slope are likely to be of low accuracy, since they depend on four parameters (the dates and magnitudes of the beginning and end of the hump) which are themselves only approximate.

The dimmest recorded hump started at magnitude 9.3, and humps can begin at any magnitude brighter than this. Since all recorded minima are dimmer than 9.3, this means that there is always a gap of at least 1.2 magnitudes between minimum and hump and, consequently, always some time delay between minimum and hump onset.

The defining characteristic of this limit seems to be the magnitude itself, not the gap, since the gap size is greater, in general, when the hump follows a dim minimum.

Beyond this, humps can appear at any brightness, right up to the approximate maximum of the star, with an apparent preference for magnitudes close to 8. See Figure 5.

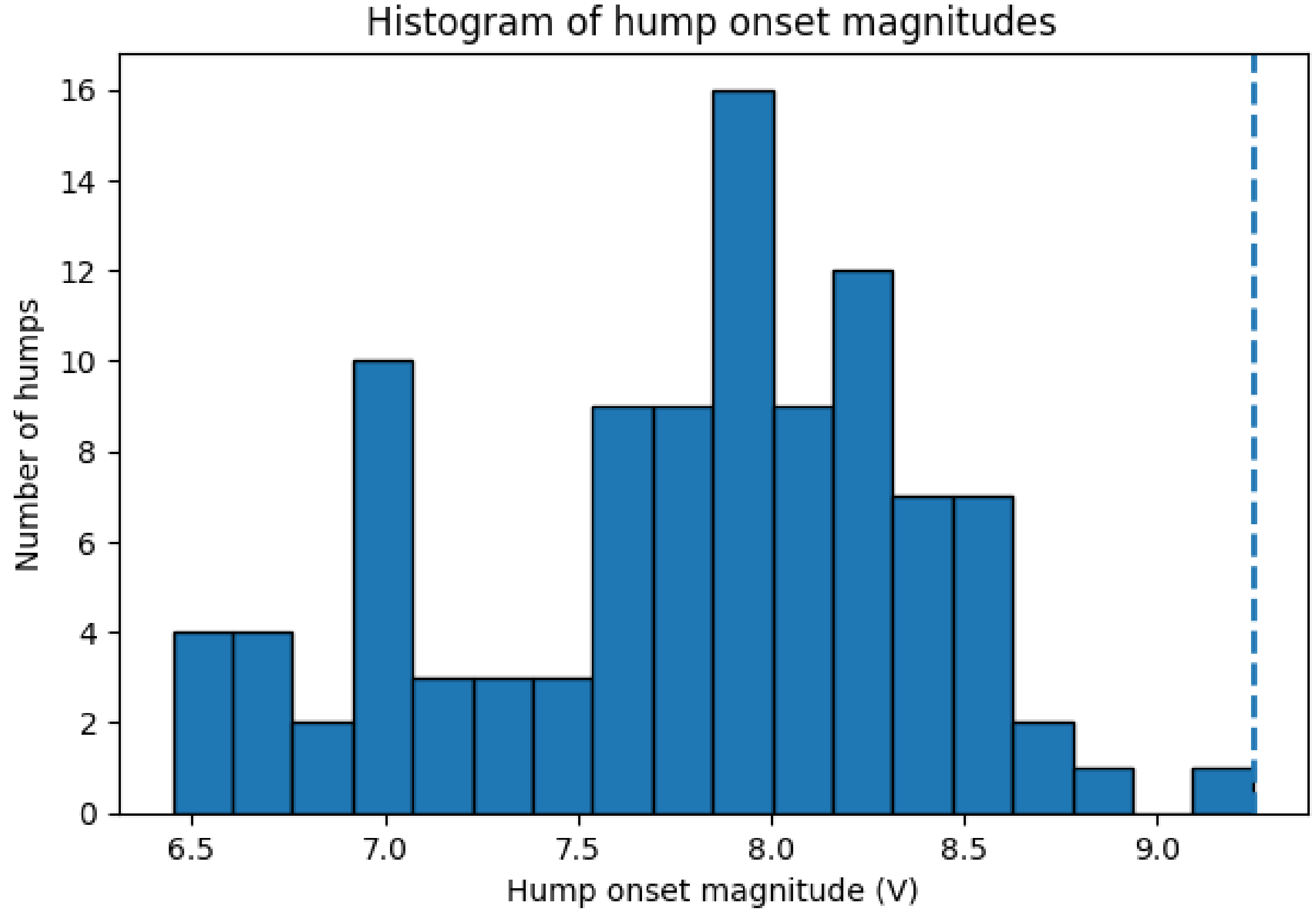

**Figure 5. Magnitudes at which humps appear**

The AAVSO light curve shows that over the 145-year observational history of T Cep there have been slow changes in both amplitude and period (Figures 6, 7 and 8). Such long-term fluctuations are common among Mira variables (Percy & Colivas 1999; Smith 2013; Merchan-Benitez et al. 2023). In the case of T Cep, there seems no reason to suspect that these changes are either periodic or evolutionary.

The amplitude change is apparently due to changes in magnitudes of both maxima and minima – there is no sign of a definite variation in mean light.

A contrast between the behaviours of period and amplitude is that, while the minimum of the period trend is at about Cycle 78, the minimum of the amplitude trend is around Cycle 88.

Comparing Figures 7 and 8 shows a marked divergence in behaviour of maximum timings and minimum timings in recent cycles, to be discussed in the next section.

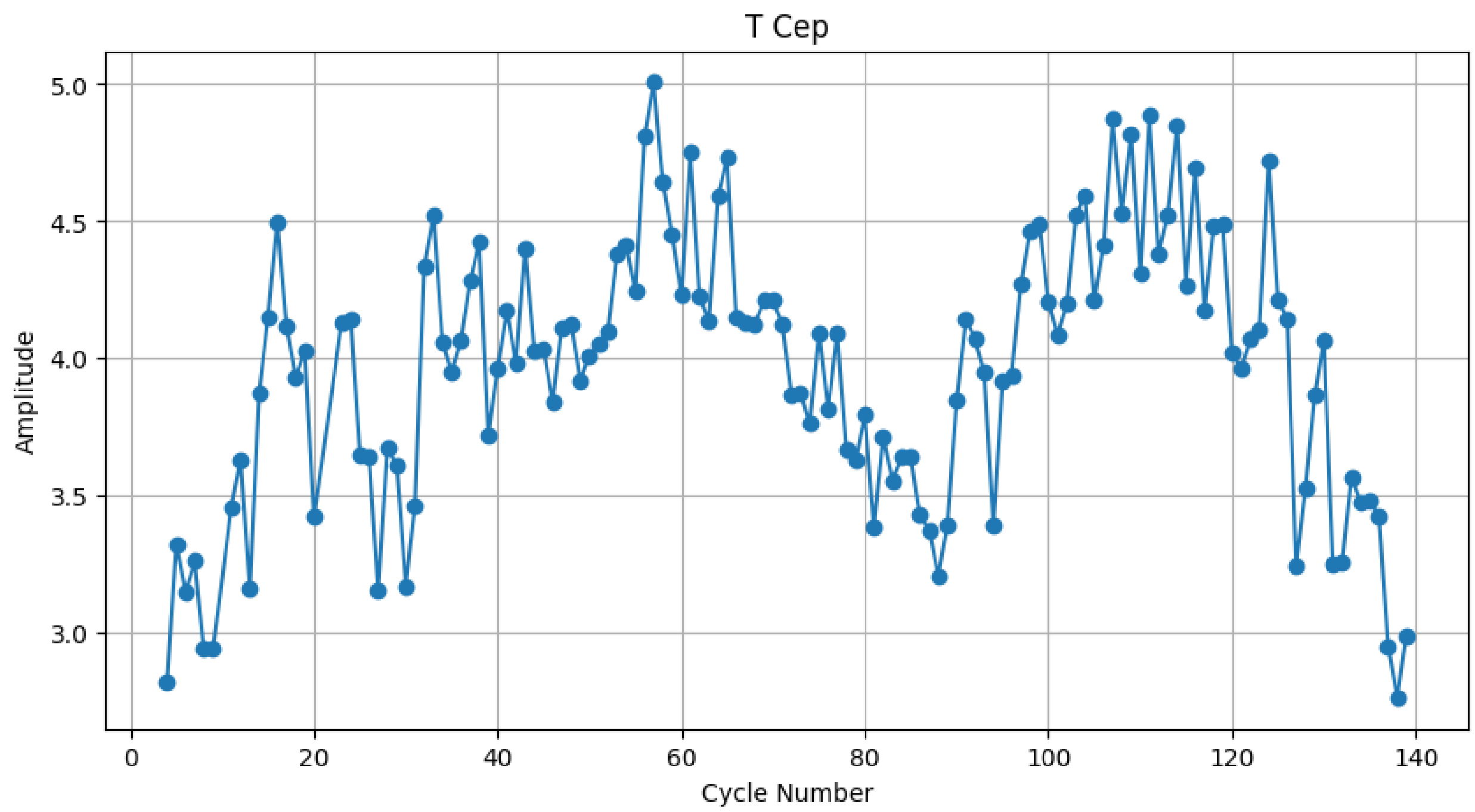

**Figure 6. Amplitudes since discovery**

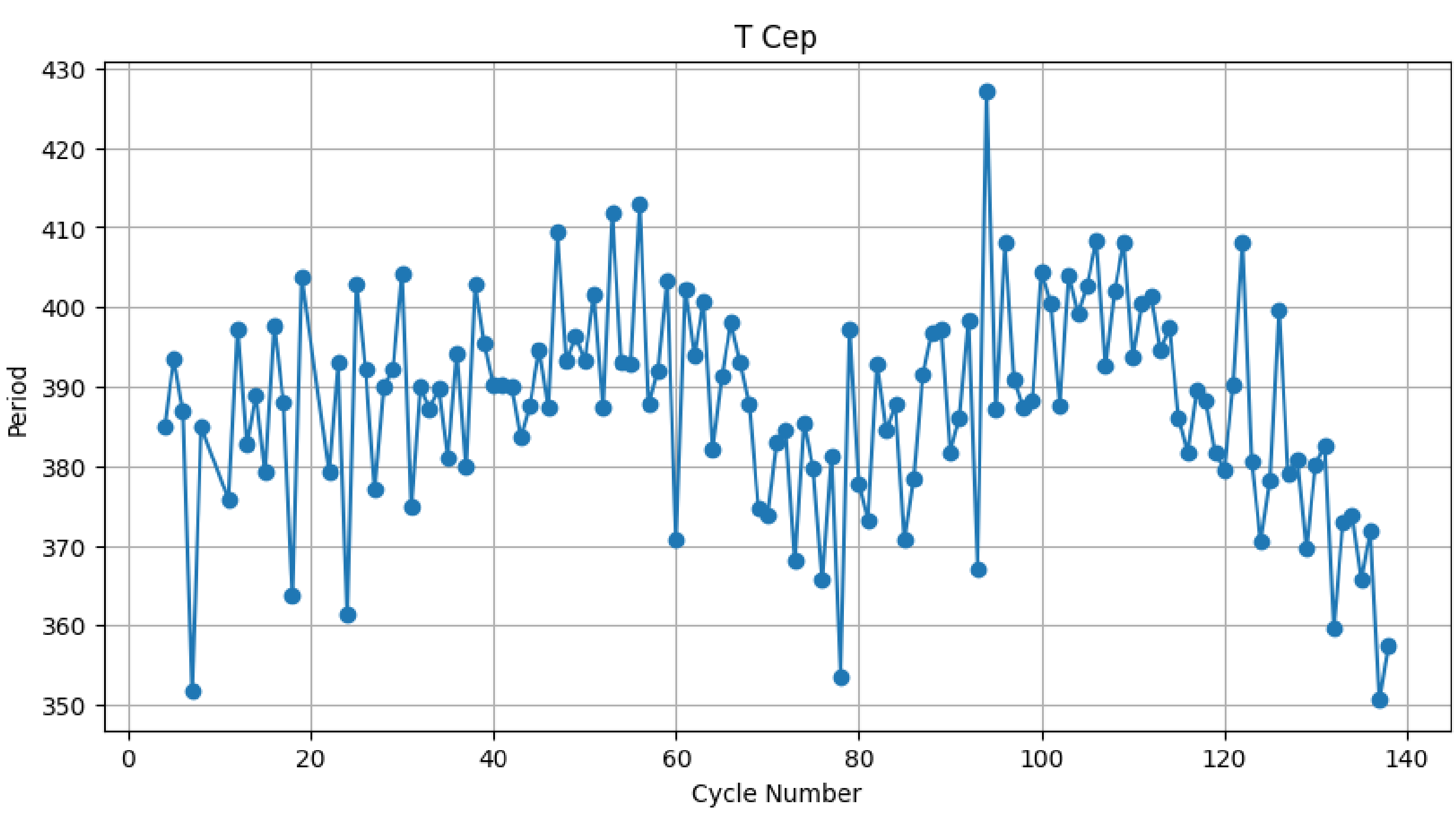

**Figure 7. Maximum-to-maximum periods**

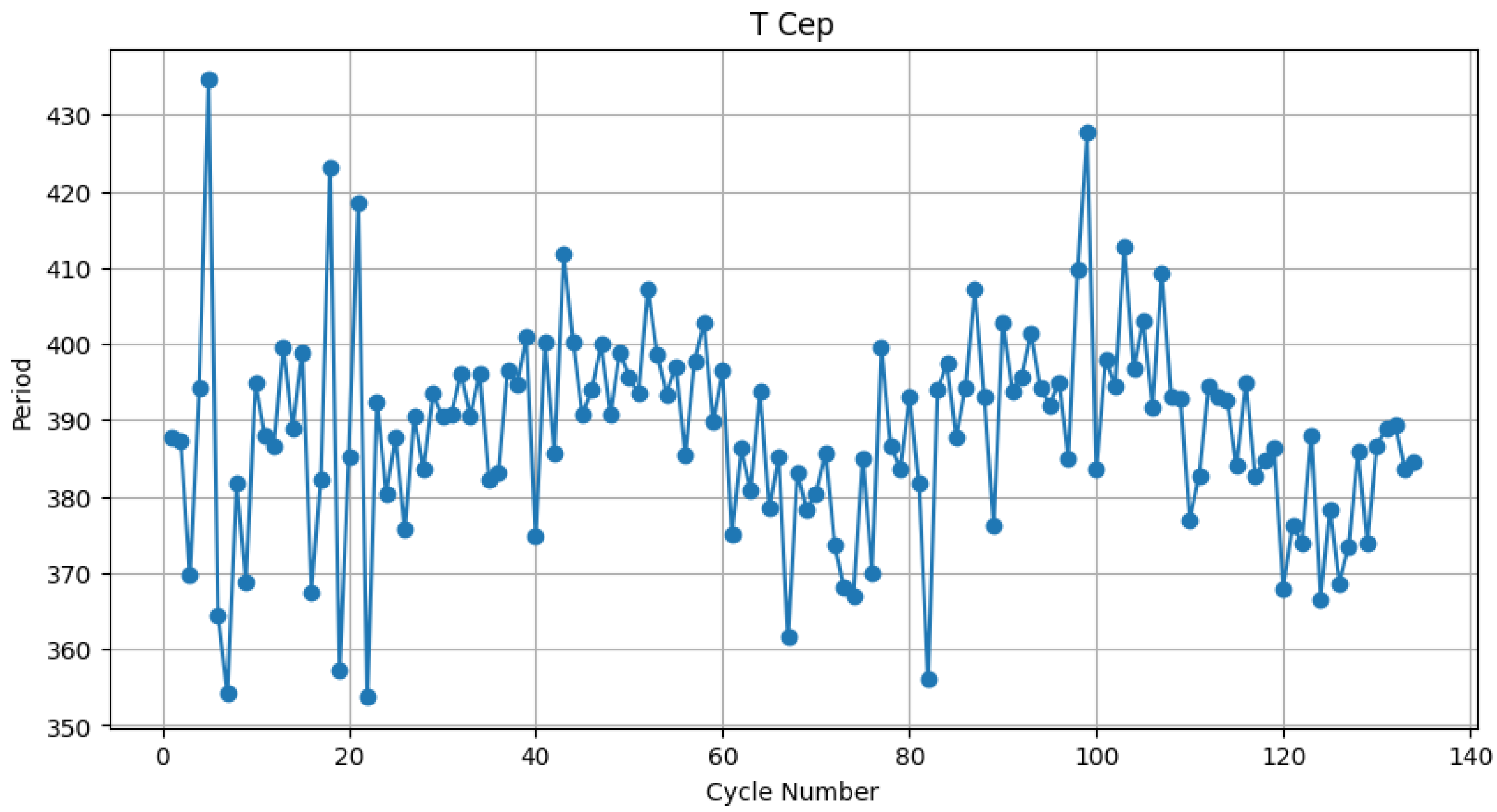

**Figure 8. Minimum-to-minimum periods**

***Recent changes***

As Figure 9 shows, over the last few years the light curve of T Cep has undergone a significant change, the most noticeable feature of which is the appearance of a hump on the descending branch of Cycle 139 (Cycles 137 and 138 also show disturbances on their descending branches which might indicate small humps).

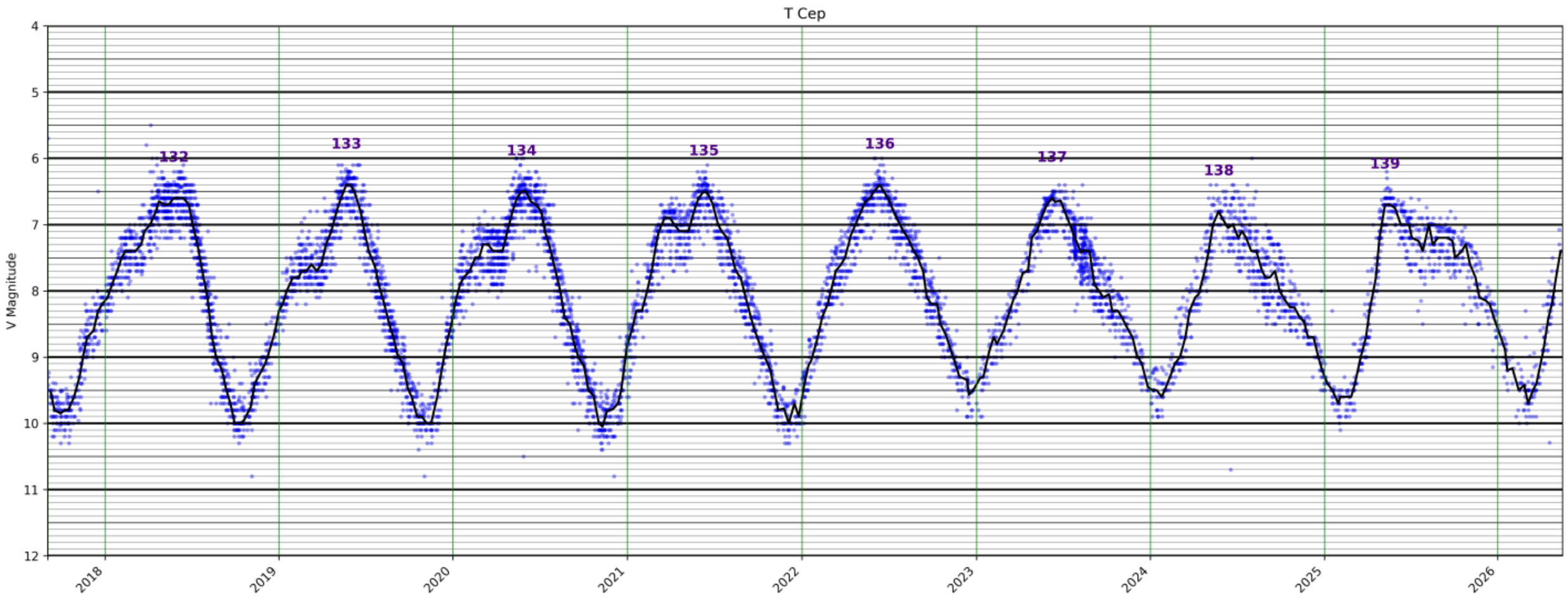

**Figure 9. Recent cycles, showing switch of hump to descending branch. The black line is the median V magnitude in successive 10-day bins of the observations, with each plotted point on the line placed at the median Julian Date of the observations within that bin.**

A hump on the descending branch seems never before to have been observed for T Cep. Humps on this branch have occasionally been noted for other Mira stars but are extremely rare and usually very slight. Hoai et al. (2025) mention SV Cassiopeiae as having an exceptional hump on the descending branch, and R Hydrae and R Serpentis have irregularities on their descending branches which may sometimes amount to humps (see for example figure 10).

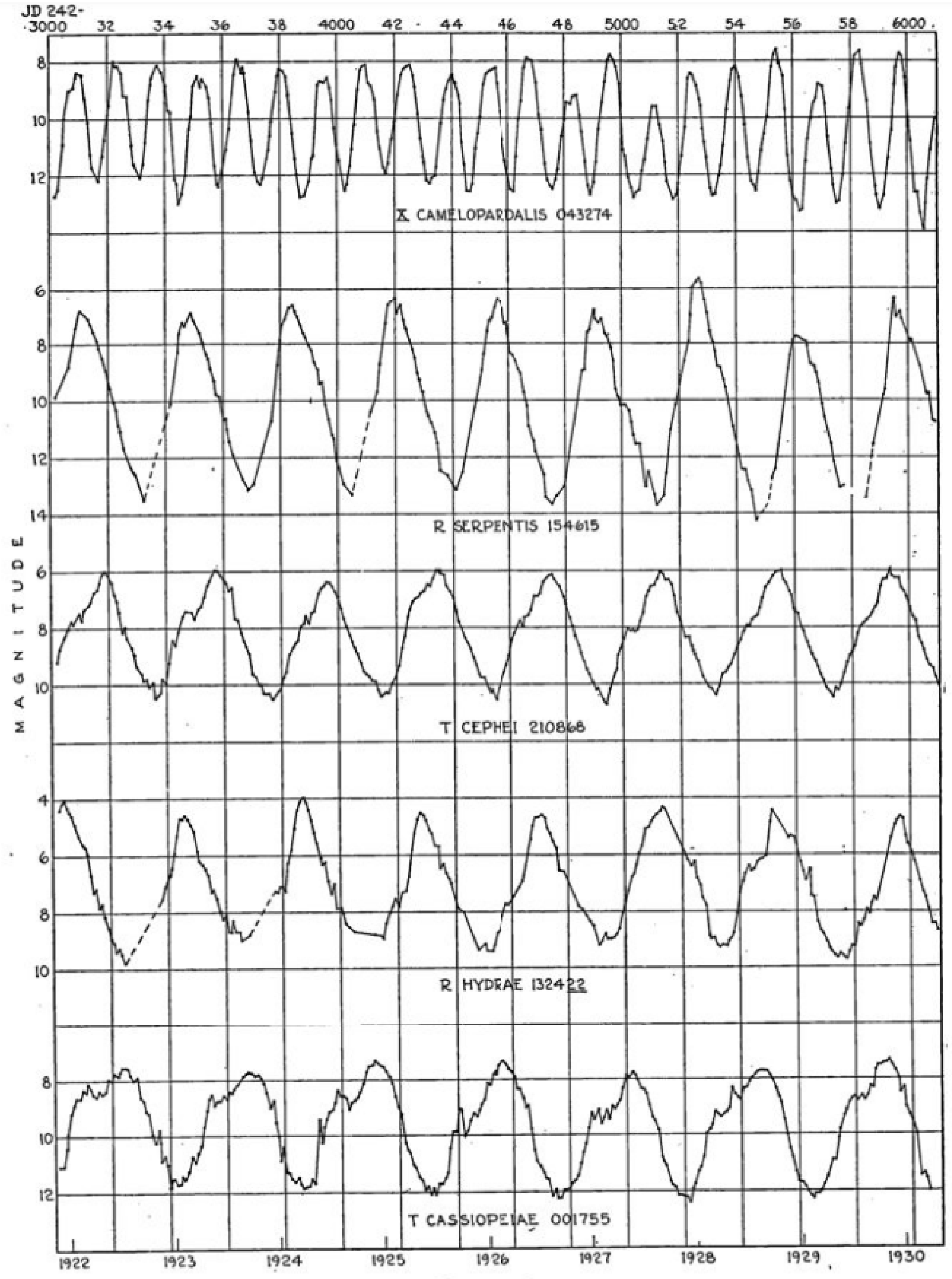

**Figure 10. Light curves of Long Period Variables. Image reproduced from Merrill (1931). Public-domain status believed likely due to age.**

T Cep's most recent cycles have included two other unprecedented features:

Cycle 138 has the smallest amplitude ever observed, at 2.8 magnitudes, and the amplitudes of Cycles 137 (2.9) and 139 (3.0) have also been much lower than the 145-year historical average of 4.0.

The maximum-to-maximum period from Cycle 137 to 138 has also been the smallest ever, at 351 days, compared to a 388 day average. However, the minimum-to-minimum period from Cycle 137 to 138 was 385 days, close to the historical average of 388. In other words, the cycles have become more asymmetric. Defining asymmetry as

$$\frac{Duration\ of\ rising\ branch - Duration\ of\ falling\ branch}{Duration\ of\ rising\ branch + Duration\ of\ falling\ branch}$$

and plotting this as Figure 11 shows this recent change clearly: throughout most of the observing history of T Cep, cycles were slightly asymmetric, with longer-lasting ascending branches than descending ones (about 0.04 on average, up to and including Cycle 126). Then, from about Cycle 127, the maxima became somewhat delayed (asymmetry became more positive) and then, perhaps from Cycle 132, this trend reversed, with maxima appearing progressively earlier in each cycle up to the most recent, highly asymmetric, Cycle 139.

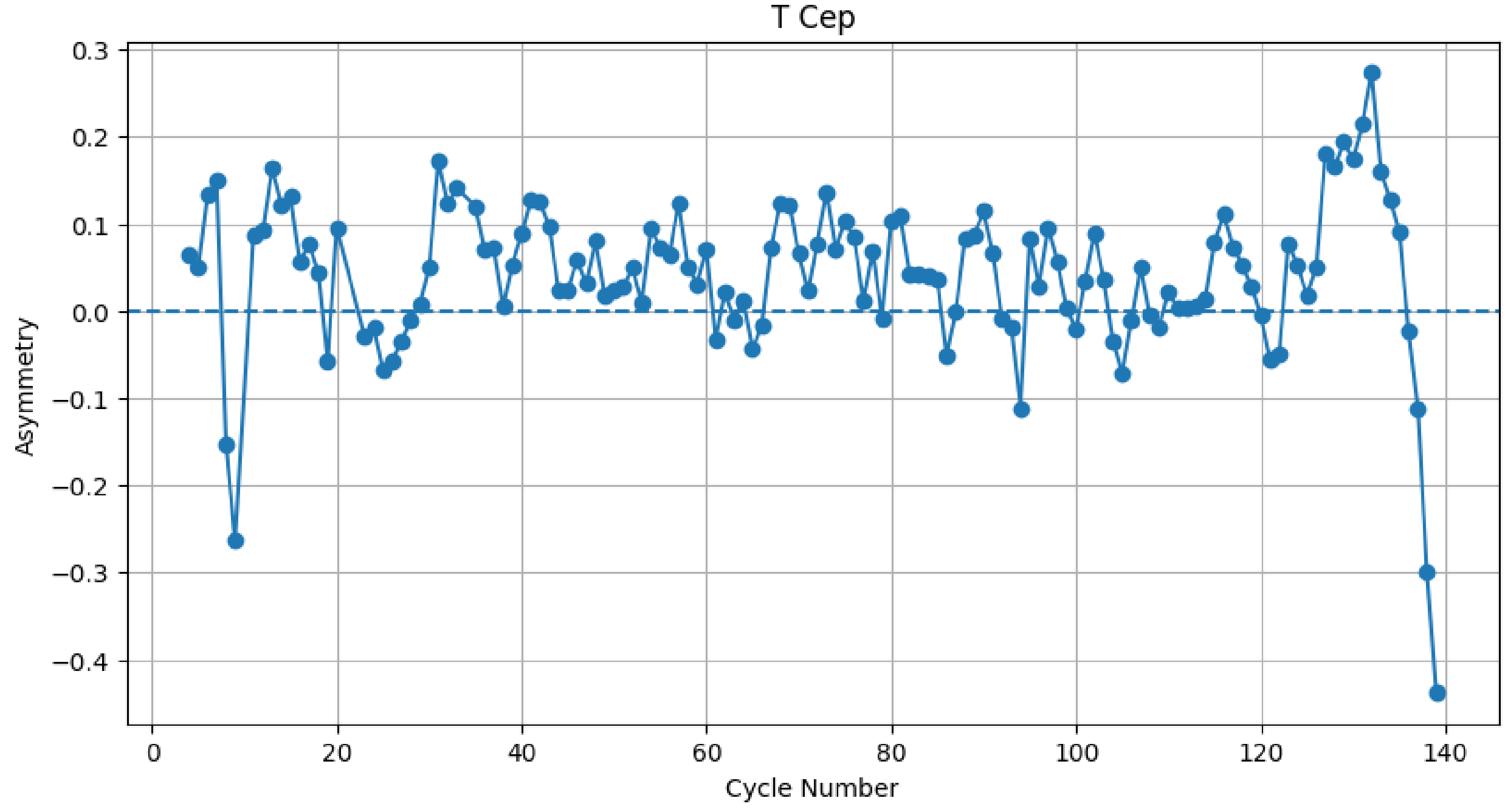

**Figure 11. Progress of asymmetries**

This increase in asymmetry contrasts the current behaviour of T Cep with the last time the amplitude and period were very low, between about Cycles 78 and 88. At that time, the asymmetries were not unusual. Conversely, the late nineteenth-century Cycles 7 and 8, which were also of low amplitude and period *do* show large asymmetries but apparently do not have humps on the declining branch (though observations at that time were relatively sparse and maxima and minima data may therefore be inaccurate - the maximum of Cycle 8 in particular is represented by a single observation – or, conversely, humps may have been missed. See Figure 12).

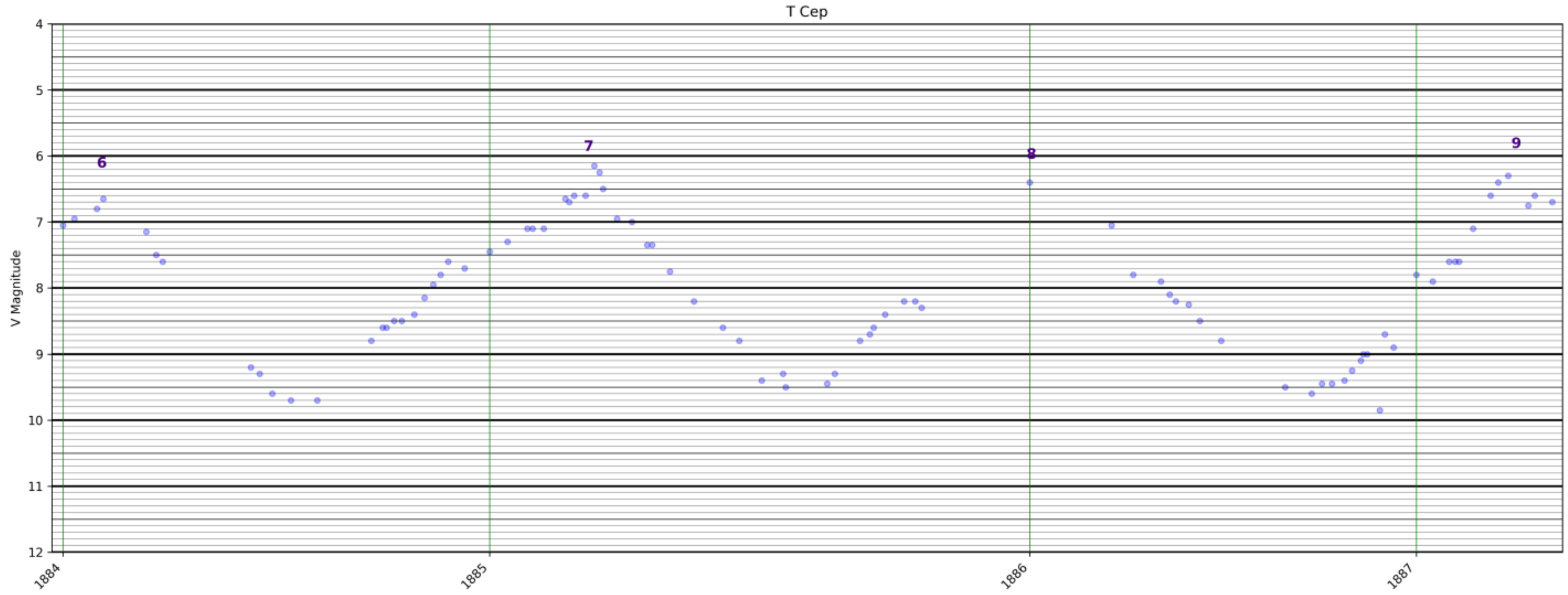

**Figure 12. Early cycles**

Accompanying this change in symmetry there may have been a progression to the hump shift for the last few cycles: in Cycle 138, there is a hint of a hump on both branches, though neither are definitive. In fact there have been no clear humps on either branch for the last few cycles. However, there is no indication that the humps of recent cycles have become progressively later.

Since the minimum period has declined with the star's amplitude, while the maximum period has recently reversed its trend by starting to shorten, it seems that it is the mechanism that determines the timing of the maxima that is connected with the shifting of the hump.

## Discussion

*Maxima and minima*

In common with other Miras, there have been slow drifts in both period and amplitude since observations began. Since the amplitude shifts seem to lag the period changes, it may be that the period alteration represents a change in pulsation frequency, while the amplitude variation is due to consequential changes in the star's outer envelope.

With the possible exception of some early cycles and the definite exception of most recent ones, observed maxima/minima have always been slightly asymmetric, with a tendency for the star to brighten more slowly than it dims.

*Humps*

The amplitudes at which humps appear seem to have four characteristics.

1. For a hump to appear, the star must exceed a brightness of about magnitude 9.3. It may be that this corresponds to an energy threshold below which the temperature, ionisation, opacity structure, or shock conditions in the atmosphere are unsuitable.
2. The cycle minimum depth apparently influences how soon the hump forms after crossing the above threshold: cycles with deeper minima tend to have dimmer humps (correlation is ~ 0.61). Possibly the lower temperatures associated with dimmer minima allow TiO formation to take place sooner, or conversely, since deep minima are an aspect of large amplitudes, perhaps enhanced pulsation energy is a factor (though correlation between amplitude and hump start brightness is somewhat weaker, at ~ -0.44[4]).
3. Dimmer humps tend to slope up more steeply – that is, their effect on the light-curve is less conspicuous. This might imply either that the hump material is relatively optically thin and therefore does not greatly obscure the photospheric brightening, or perhaps that the material is slower to form in such humps.
4. Notwithstanding 2. above, the majority of humps appear at around magnitude 8.

The fact that hump presence, brightness and duration are all correlated from cycle to cycle implies that the atmospheric conditions that influence these parameters persist on longer timescales than the pulsation cycle.

It seems that different conditions influence hump brightness and hump duration, since these are not correlated with each other: whatever the trigger condition for humps is, it is different to the condition which sustains them.

The question arises as to whether a hump is caused by some condition – such as a dense layer of TiO molecules - which forms suddenly, persists for the duration of the hump, and is then abruptly dissipated. One could then explain the upward slopes of some humps as caused by optically thin but persistent TiO layers, through which the brightening of the photosphere can be seen.

However, some humps clearly do slope downward (such as that of Cycle 73 in Figure 13), so they must increase in opacity while the photosphere behind them continues to brighten, showing that, in some cases at least, the obscuring material increases through the lifetime of the hump before being suddenly removed.

[4] The change of sign arises since dimmer magnitudes are numerically larger.

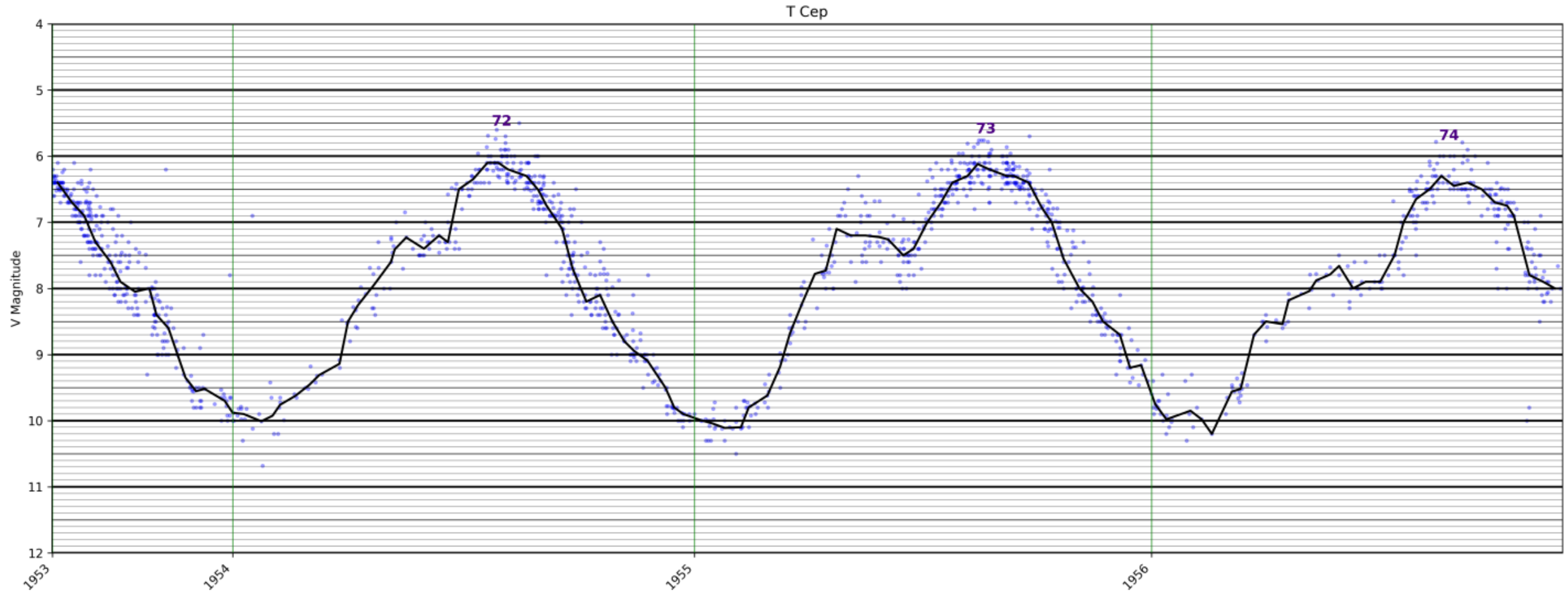

**Figure 13. Example of a down-sloping hump**

This in turn implies that a hump comes to an end, not because a gradual process of destruction has run its course by then, but because the conditions which permit it to exist have suddenly changed.

If the hump is indeed caused mainly by increased opacity due to TiO formation, then the excess moelcules must largely dissipate within each cycle otherwise there would be a long-term build up of unprocessed TiO which would cause a gradual dimming of the star.

One tentative model is that T Cep sometimes generates additional transient pressure pulses which, while passing through a layer rich in titanium and oxygen, increase the local pressure so that TiO formation – and hence opacity - is enhanced, thereby causing a hump. Humps may slope upward or downward depending on the balance between TiO formation, TiO destruction, and the underlying brightening rate of the photosphere. Once the pulse passes the upper boundary of this layer, the ambient pressure falls, allowing molecules to dissociate rapidly, reducing the opacity and bringing the hump to an abrupt end. The approximately constant molecular abundance during the hump would explain the unchanging TiO spectral features in Figure 2. The fact that the Ca II triplet brightens throughout the hump supports the idea that the hump is local to the outer atmosphere: the photosphere below it continues brightening throughout the ascending branch whether or not a visual hump is present.

According to this model, the duration of the hump depends on the time for the pressure wave(s) to traverse the TiO region. Assuming that the wave speed does not change greatly from cycle to cycle, the implication is that the extent of the region changes over a period of a few cycles.

Kudashkina & Rudnitskij (1994) suggested that humps on the ascending branches of Mira light curves may be associated with the propagation of shock waves through stellar atmospheres, noting that the durations of such features are comparable to estimated atmospheric shock-crossing times.

The interpretation proposed here differs in that the hump phenomenon is treated as a transient opacity event associated with a secondary localized compressive pulse rather than with the primary pulsation shock itself.

These secondary pressure pulse systems, if they exist, must form somewhat independently of the regular shock sequences which cause the normal pulsation of the star, since not all cycles have humps, and hump formation occurs at a range of star brightnesses.

Nevertheless, when pulsation amplitude is high one might expect some increase in secondary pulsation strength, which could explain why hump formation is more likely to occur in such cycles: the more energetic waves involved could force molecule formation at deeper, hotter levels.

The observed hump durations in T Cep range from 10-111 days. The following estimate checks whether this timescale is compatible with the time required for a transient compressive pulse, moving at the local sound speed, to cross a TiO-forming region in the outer atmosphere.

The sound speed in a gas is approximately:

$$c_s = \sqrt{\frac{\gamma kT}{\mu m_H}}$$

where $\gamma$ is the adiabatic index, $k$ the Boltzmann constant, $T$ the absolute temperature, $\mu$ the mean molecular weight, and $m_H$ the mass of a hydrogen atom. For temperatures of 2000 to 3000 K, this implies speeds of 3 to 6 km/s.

If the duration of a hump represents the time $t$ taken for such a pulse to traverse the relevant TiO-rich region, then the corresponding path length is:

$$L = c_s t$$

So, a 10-day hump implies to a crossing distance of about 0.02 - 0.04 AU, while a 111-day hump would be about 0.2 - 0.4 AU. For T Cep these figures correspond to about 1–30% of the stellar radius.

This is significantly smaller than the TiO-rich layer of Mira’s atmosphere, which extends to about 4 stellar radii (Kamiński, et al. 2017) and is presumably not dissimilar to that of T Cep. Hence the observed 10-111 day hump durations are not obviously inconsistent with sound-speed crossing times through atmospheric regions in which TiO formation could occur.

This calculation should not be read as evidence that the proposed mechanism is correct. It only shows that the timescale is physically plausible. The proposed model remains highly speculative in the absence of spectroscopic evidence to support it.

*Recent changes*

The recent changes in T Cep's light curve seem somewhat similar to those observed in the Mira variable T Ursae Minoris (Molnár et al. 2019). In the late twentieth and early twenty-first centuries, T UMi underwent a major long-term decline in both pulsation period and visual amplitude, with the period decreasing from approximately 315 days to about 200 days and the amplitude falling from roughly 4–5 magnitudes to around 1.5–2 magnitudes. During this transition the light curve became increasingly complex and developed secondary structures, including humps on the descending branch. See Figure 14.

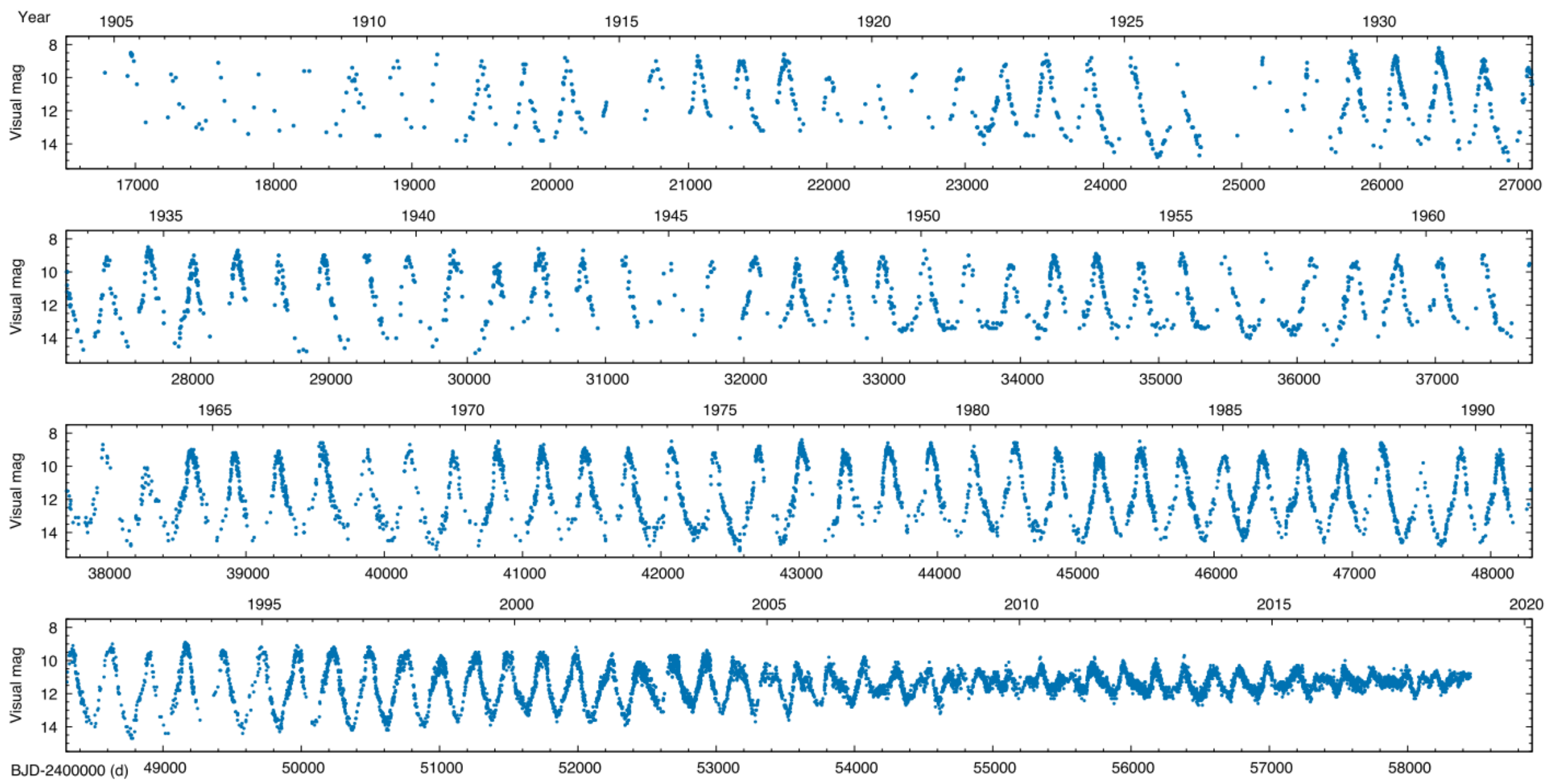

**Figure 14. AAVSO light curve of T Umi, from (Molnár et al. 2019), by kind permission of László Molnár**

Molnár et al. conclude that the T Umi is undergoing evolutionary developments; the immediate cause of the radical light-curve changes being a thermal pulse (helium-shell flash), producing contraction of the stellar envelope and substantial changes in the pulsation properties of the star.

The recent behaviour of T Cep shows several qualitative similarities: a long-term reduction in both period and amplitude, together with a major alteration in light-curve morphology which includes a descending-branch hump.

Hence, it is possible that the recent change in T Cep could also be evolutionary in nature.

Whether or not this is the case, one possible explanation of the declining-branch hump is that, as a result of gradually decreasing pulsational energy over recent cycles (evidenced by the observed reductions in amplitude), atmospheric temperatures have dropped sufficiently to allow widespread formation of TiO and perhaps other opacity-producing molecules.

Whether T Cephei really is undergoing evolutionary change will become clearer over the next few cycles. Spectroscopic observations during those cycles, especially at different stages of any future humps, should permit much firmer conclusions regarding the atmospheric dynamics responsible for the complexities of its light curve.

*I gratefully acknowledge the contributions of the AAVSO observer community, whose photometric data and metadata resources were used in this study and made available through the AAVSO's scientific archives.*

**References**

Boyd, D. 2023, "Spectroscopic and Photometric Study of the Asymptotic Giant Branch Star T Cephei", *Journal of the American Association of Variable Star Observers*, 51, 220–237

Ceraski, W. 1879, "Über einen neuen Veränderlichen", *Astronomische Nachrichten*, 94, 175.

Gavin, M. 1999, "Pre-maximum Spectra of T Cep", British Astronomical Association Variable Star Section. Available at: BAA VSS T Cep spectra page (accessed 25 May 2026).

Hoai, D. T., Nhung, P. T., Darriulat, P., & Tan, M. N. 2025, "From the Light Curves of Long Period Variables to their Evolution along the AGB", *The Astrophysical Journal*, submitted.

Kamiński, T., Müller, H. S. P., Schmidt, M. R., Cherchneff, I., Wong, K. T., Brünken, S., Menten, K. M., Winters, J. M., Gottlieb, C. A., & Patel, N. A. 2017, "An observational study of dust nucleation in Mira (o Ceti): II. Titanium oxides are negligible for nucleation at high temperatures", *Astronomy & Astrophysics*, 599, A59.

Kudashkina, L. S., & Rudnitskij, G. M. 1994, "Influence of Shock Waves on the Light Curves of Long-Period Variables", *Odessa Astronomical Publications*, 7, 66.

Lebzelter, T. 2011, "The Shapes of Light Curves of Mira-Type Variables", *Astronomische Nachrichten*, 332, 140–148.

Merchan-Benitez, P., Uttenthaler, S., & Jurado-Vargas, M. 2023, “Long-term Variability of Mira Stars from Historical and Modern Surveys”, *Astronomy & Astrophysics*, 672, A54.

Merrill, P. W. 1931, “Light Curves of Long Period Variables”, *Popular Astronomy*, 39, 121.

Molnár, L., et al. 2019, “Stellar Evolution in Real Time. I. Models Consistent with the Direct Observation of a Thermal Pulse in T Ursae Minoris”, *The Astrophysical Journal*, 879, 62.

Percy, J. R., & Colivas, T. 1999, “Long-term Changes in Mira Stars”, *Publications of the Astronomical Society of the Pacific*, 111, 94.

Smith, H. A. 2013, “Long-term Behaviour of Mira Variables”, *Journal of the American Association of Variable Star Observers*, 41, 475.

Templeton, M. R., & Willson, L. A. 2004, “Secular Evolution in Mira Variables”, *American Astronomical Society Meeting Abstracts*, 205, 54.07.

Vardya, M. S. 1988, “Classification of Mira Variables Based on Visual Light Curve Characteristics”, *Astronomy & Astrophysics Supplement Series*, 73, 181–194.